\documentclass{aastex61}
\usepackage{amsmath}
\usepackage[]{hyperref}
\usepackage{longtable}

\shorttitle{CTA 102 imaging and variability}
\shortauthors{Li et al.}

\begin{document}

\title{Imaging and variability studies of CTA~102 during the 2016 January $\gamma$-ray flare}

\correspondingauthor{P. Mohan}
\email{pmohan@shao.ac.cn}
\correspondingauthor{T. An}
\email{antao@shao.ac.cn}
\author{Xiaofeng Li}
\affiliation{Shanghai Astronomical Observatory, 80 Nandan Road, Shanghai 200030, China}
\affiliation{Shanghai Tech University, 100 Haike Road, Pudong, Shanghai, 201210, China}
\affiliation{University of Chinese Academy of Sciences, 19 A Yuquan Rd, Shijingshan District, Beijing 100049, China   }
\author{P. Mohan}
\affiliation{Shanghai Astronomical Observatory, 80 Nandan Road, Shanghai 200030, China}
\author{T. An}
\affiliation{Shanghai Astronomical Observatory, 80 Nandan Road, Shanghai 200030, China}
\affiliation{Key Laboratory of Radio Astronomy, Chinese Academy of Sciences, 210008 Nanjing, P.R. China}
\author{Xiaoyu Hong}
\affiliation{Shanghai Astronomical Observatory, 80 Nandan Road, Shanghai 200030, China}
\affiliation{Shanghai Tech University, 100 Haike Road, Pudong, Shanghai, 201210, China}
\affiliation{Key Laboratory of Radio Astronomy, Chinese Academy of Sciences, 210008 Nanjing, P.R. China}
\author{Xiaopeng Cheng}
\affiliation{Shanghai Astronomical Observatory, 80 Nandan Road, Shanghai 200030, China}
\affiliation{University of Chinese Academy of Sciences, 19 A Yuquan Rd, Shijingshan District, Beijing 100049, China   }
\author{Jun Yang}
\affiliation{Department of Earth and Space Sciences, Chalmers University of Technology, Onsala Space Observatory, SE-439 92 Onsala, Sweden}
\author{Yingkang Zhang}
\affiliation{Shanghai Astronomical Observatory, 80 Nandan Road, Shanghai 200030, China}
\affiliation{University of Chinese Academy of Sciences, 19 A Yuquan Rd, Shijingshan District, Beijing 100049, China   }
\author{Zhongli Zhang}
\affiliation{Shanghai Astronomical Observatory, 80 Nandan Road, Shanghai 200030, China}
\affiliation{Key Laboratory of Radio Astronomy, Chinese Academy of Sciences, 210008 Nanjing, P.R. China}
\author{Wei Zhao}
\affiliation{Shanghai Astronomical Observatory, 80 Nandan Road, Shanghai 200030, China}
\affiliation{Key Laboratory of Radio Astronomy, Chinese Academy of Sciences, 210008 Nanjing, P.R. China}


\begin{abstract}
The $\gamma$-ray bright blazar CTA 102 is studied using imaging (new 15 GHz and archival 43 GHz Very Long Baseline Array, VLBA data) and time variable optical flux density, polarization degree and electric vector position angle (EVPA) spanning between 2015 June 1 and 2016 October 1, covering a prominent $\gamma$-ray flare during 2016 January. The pc-scale jet indicates expansion with oscillatory features upto 17 mas. Component proper motions are in the range $0.04 - 0.33$ mas yr$^{-1}$ with acceleration upto 1.2 mas followed by a slowing down beyond 1.5 mas. A jet bulk Lorentz factor $\geq 17.5$, position angle of $128\fdg3$, inclination angle $\leq 6\fdg6$ and intrinsic half opening angle $\leq 1\fdg8$ are derived from the VLBA data. These inferences are employed in a helical jet model to infer long term variability in flux density, polarization degree, EVPA and a rotation of the Stokes Q and U parameters. A core distance of $r_{\rm core,\,43 \ GHz}$ = 22.9 pc, and a magnetic field strength at 1 pc and the core location of $1.57$ G and $0.07$ G respectively are inferred using the core shift method. The study is useful in the context of estimating jet parameters and in offering clues to distinguish mechanisms responsible for variability over different timescales.
\end{abstract}


\section{Introduction} \label{sec:intro}

CTA~102 \cite[2230+114; at redshift of 1.037: ][]{1965ApJ...141.1295S} is a high-polarization \citep{1981ApJ...243...60M} and optically violent variable quasar \citep{1986ApJ...310..325M}. It was identified as a flat-spectrum radio-loud quasar (FSRQ; a sub-class of blazars) based on a $1 - 22$ GHz spectral index of $\alpha \geq -0.5$ ($S \propto \nu^{\alpha}$) obtained from RATAN-600 telescope observations \citep{1999A&AS..139..545K,2004ApJ...609..539K}. It is a prominent $\gamma$-ray source detected with the Energetic Gamma-Ray Experiment Telescope (EGRET) on the Compton Gamma Ray Observatory \cite[e.g.][]{1993ApJ...414...82N, 1994ApJS...94..551F} and also by the {\it Fermi}/LAT \cite[e.g., ][]{2009ApJ...700..597A}.

Early 15-GHz images using the very long baseline interferometry (VLBI) technique obtained from the very long baseline array (VLBA) indicated a twisted morphology with jet bending on a scale of $\sim$20 mas \citep{1998AJ....115.1295K}, and multi-epoch 43 GHz observations show jet knots with complex kinematics involving a mixture of apparent superluminal motion and as well as stationary components \citep{2001ApJS..134..181J,2005AJ....130.1418J}. Multi-frequency (15 and 43 GHz) multi-epoch observations during the 2006 radio flare \citep{2013A&A...551A..32F} inferred a possible association between a jet component ejection event at the end of 2005 and a strong radio flare in 2006 April. The authors interpreted the 2006 radio flare as a result of the interaction between a propagating shock and a stationary shock at a de-projected distance of 18 pc from the core. Temporal variability studies during the 2012 September-October multi-band flaring period found a near-simultaneous $\gamma$-ray and optical flaring behaviour, inferring a co-spatial origin \citep{2014ApJ...797..137C,2016MNRAS.461.3047L}. The latter study in addition suggested that the measured Stokes parameter variations is consistent with a bright jet knot moving along a helical trajectory. A multi-wavelength polarimetric study during the same flaring phase \citep{2015ApJ...813...51C} detected a co-spatial origin from the near-simultaneous variability and identified the passage of a superluminal radio knot coincident with the $\gamma$-ray flare. Further evidence, including an intra-day optical polarization variability and clockwise rotation of the electric vector position angle (EVPA, $\chi = \frac{1}{2} \arctan \frac{U}{Q}$, where $U$ and $Q$ are Stokes components) during the flaring phase, is consistent with a jet knot passing a region hosting helical magnetic fields. A study of $\gamma$-ray -- optical variability in flux density and polarization between prominent flares during the end of 2016 \citep{2017Galax...5...91L} finds no time lag between the light curves indicating co-spatial origin of synchrotron (optical) and inverse-Compton ($\gamma$-ray) flux, and a smaller viewing angle (more energetic, emission closer to jet base) compared to a flare in 2012 when interpreted in terms of a blob or shock wave on a helical trajectory. From the above studies, the $\gamma$-ray flares may be associated with optical flux and polarization variability, and the jet kinematics and polarization properties may be described in terms of a helical jet. 

The current work is aimed at inferring jet properties using imaging and timing data (optical flux density, polarization degree and EVPA) spanning $\sim$ 1 year. The source was imaged during a large $\gamma$-ray flare in 2016-January, thus providing an excellent opportunity to investigate the jet properties during this event. 
In Section \ref{sec:obs}, we present the observation data ($\gamma$-ray; optical R-band photometry, polarization and EVPA; 15 and 43 GHz VLBI radio images) and data reduction. The results from this study are given in Section \ref{sec:results}, including the pc-scale jet morphology and polarization structure, component kinematics and jet parameters. A helical jet model and the application of the core shift effect to derive the jet parameters are presented in Section \ref{sec:discussion}. A summary is given in Section \ref{sec:summary}. Assuming cosmological parameters of H$_{0}$ = 71 km s$^{-1}$ Mpc$^{-1}$, $\Omega_{m} = 0.27$, $\Omega_{\Lambda} = 0.73$, an angular size of 1 mas corresponds to a projected linear length of 8.11 pc at the redshift of CTA 102 \citep{2006PASP..118.1711W}. With these conventions, 1 mas yr$^{-1}$ proper motion speed corresponds to 52.9 $c$. The current is aimed at inferring jet properties using imaging and timing data (optical flux density, polarization degree and EVPA) spanning $\sim$ 1 year. The source was imaged during a large $\gamma$-ray flare in 2016-January), thus providing an excellent opportunity to investigate the jet properties during this flare.

\section{Observations \& Data Reduction} \label{sec:obs}
\subsection{\texorpdfstring{$\gamma$}--ray and optical data}
\label{subsec:gamma}

\begin{figure}
\centerline{\includegraphics[scale=0.8]{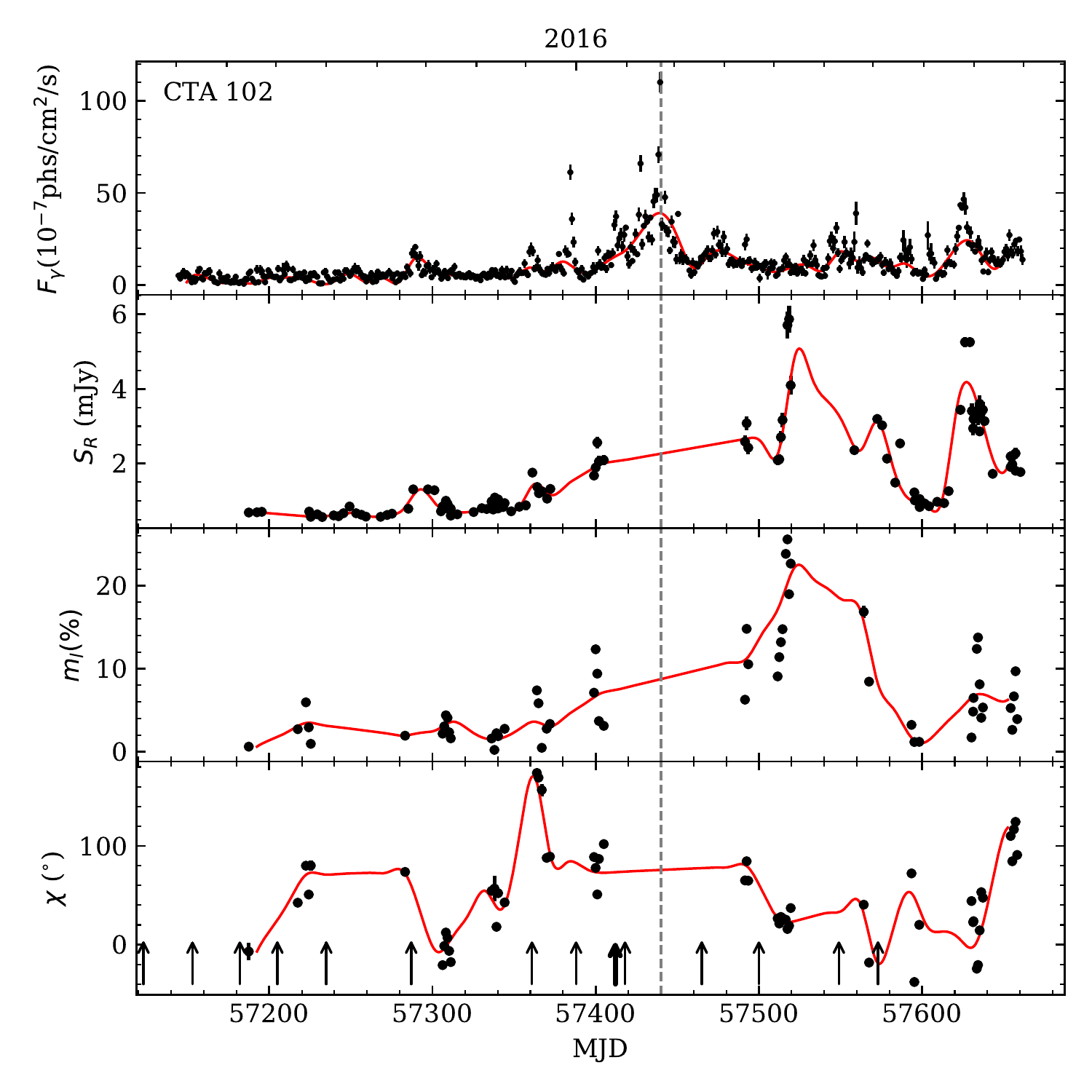}}
\caption{
The upper panel is the $\gamma$-ray light curve, obtained from the Fermi/LAT monitored source list light curves (refer to section \ref{subsec:gamma}). The second panel is the optical R-band photometric light curve. The optical polarization degree and optical EVPAs are shown in bottom two panels. All optical data are obtained from the Steward Observatory spectropolarimetric monitoring project (refer to section \ref{subsec:gamma}). The vertical arrows in the bottom panel correspond to the epochs of VLBI observations presented in this paper. The thick arrow represent 15 GHz observation while the thin arrows represent 43 GHz. The red curves are cubic spline interpolation through the 10-day binned data.}
\label{fig:lc}
\end{figure}

CTA 102 was monitored by {\it Fermi}/LAT\footnote{\href{https://fermi.gsfc.nasa.gov/ssc/data/access/lat/msl\_lc/}{\url{https://fermi.gsfc.nasa.gov/ssc/data/access/lat/msl\_lc/}}} in the 0.1--300 GeV energy range during the 2016 January flaring phase. 
The daily binned photon flux is presented in upper panel of Figure \ref{fig:lc} to illustrate the $\gamma$-ray activity during the VLBI observation period.
 CTA 102 is among a sample of thiry three $\gamma$-ray-bright blazars monitored at the Steward Observatory\footnote{\href{http://james.as.arizona.edu/psmith/Fermi/}{\url{http://james.as.arizona.edu/~psmith/Fermi/}}}. Both optical flux spectra and linear polarization data are obtained from the monitoring observations at a density of one week every month, allowing for coordinated studies of the physical connection between $\gamma$-ray and optical radiation in blazars \citep{2009arXiv0912.3621S}. 

The optical program utilizes either the 2.3 m Bok Telescope on Kitt Peak, AZ or the 1.54 m Kuiper Telescope on Mt. Bigelow, AZ to monitor these $\gamma$-ray-bright blazars. All observations are made using the SPOL spectropolarimeter and provide a spectral coverage in the 4000--7550\AA \ optical wavelength range. 
Differential flux measurements made using stars in the same observed field \citep{2009arXiv0912.3621S} were used for the optical flux calibration. The polarimetric data are polarized flux spectra over 5000-7000\AA. CTA 102 is also a target of the $\gamma$-ray bursts optical afterglow observation program \citep{2003PASP..115..844L}. The program uses the 0.76 m Katzman Automatic Imaging Telescope (KAIT) of Lick Observatory\footnote{\href{http://herculesii.astro.berkeley.edu/kait/agn/}{\url{http://herculesii.astro.berkeley.edu/kait/agn/}}} to monitor 163 AGNs in the R-band. The data was acquired from website of the program. The KAIT observation and calibration methods are described in \citet{2003PASP..115..844L}. The accuracy of magnitudes is $\sim 5\%$. The optical light curve, linear polarization and EVPA variations are presented in the second, third and last panels of Figure \ref{fig:lc} respectively.

\subsection{VLBI observations and data reduction}
\label{subsec:vlbi}
A 15 GHz VLBA observation of CTA 102 was carried out on 2016 January 25 (code: BA113; PI: T.~An) during the prominent $\gamma$-ray flare in 2016 January. All ten VLBA telescopes participated in the observation with a total of 8 hours, in which 6 hours were spent on the target CTA~102. The data were recorded on eight basebands (IFs) with a total bandwidth of 256 MHz and a recording data rate of 2 Gbps. The observations were performed in polarization mode. A bright quasar 2145$+$067 was used as the instrumental polarization (so-called `D-term') calibrator. 
BL Lac (2200+420) was used as EVPA calibrator; 
an intrinsic EVPA of $\chi=150\degr$ was determined based on MOJAVE observations\footnote{\href{http://www.physics.purdue.edu/astro/MOJAVE/}{\url{http://www.physics.purdue.edu/astro/MOJAVE/}}} \citep{2017arXiv171107802L} at 15 GHz during 2016-01-22 which we employed to calibrate the EVPA variations in CTA 102. The data were correlated in the DiFX correlator \citep{2007PASP..119..318D} in Socorro, New Mexico, USA with a 2-second integration. Cross-handed Stokes components and 64 spectral points per IF were exported.

The calibration of the visibility data is performed using Astronomical Image Processing System (AIPS) following the standard procedure described in the AIPS Cookbook. The amplitude calibration of the visibilities is carried out using the gain curves and system temperatures measured at each station during the observations. The fringe fitting solution interval used is 0.5 minutes. Then, the gain solutions are applied to the data after which, the data is exported from AIPS into the Differential Mapping software package \cite[Difmap][]{1997ASPC..125...77S}. Several iterations of phase and amplitude self-calibration were made to calibrate the residual phase errors in order to reduce the image noise. After the self-calibration, Gaussian models consisting of emission components with circular Gaussian intensity distributions were applied to fit the visibility data to quantitatively describe the emission distribution. The model has four parameters: flux density($S$), seperation from image center ($r$), position angle (PA), and size ($d$).

CTA 102 has been a target of the monthly monitoring program of bright $\gamma$-ray blazars observed with the VLBA at 43 GHz \citep{2016Galax...4...47J} since 2007\footnote{\href{http://www.bu.edu/blazars/VLBAproject.html}{\url{http://www.bu.edu/blazars/VLBAproject.html}}}. The visibilities data between 2015 February to 2016 July were calibrated by the Boston University (BU) group. 
The observational logs of the 15 and 43 GHz VLBA observations are given in Table \ref{tab:obs}.

\section{Results} \label{sec:results}

\subsection{Imaging and pc-scale morphology}

\begin{figure}
\centerline{\includegraphics[width=\linewidth]{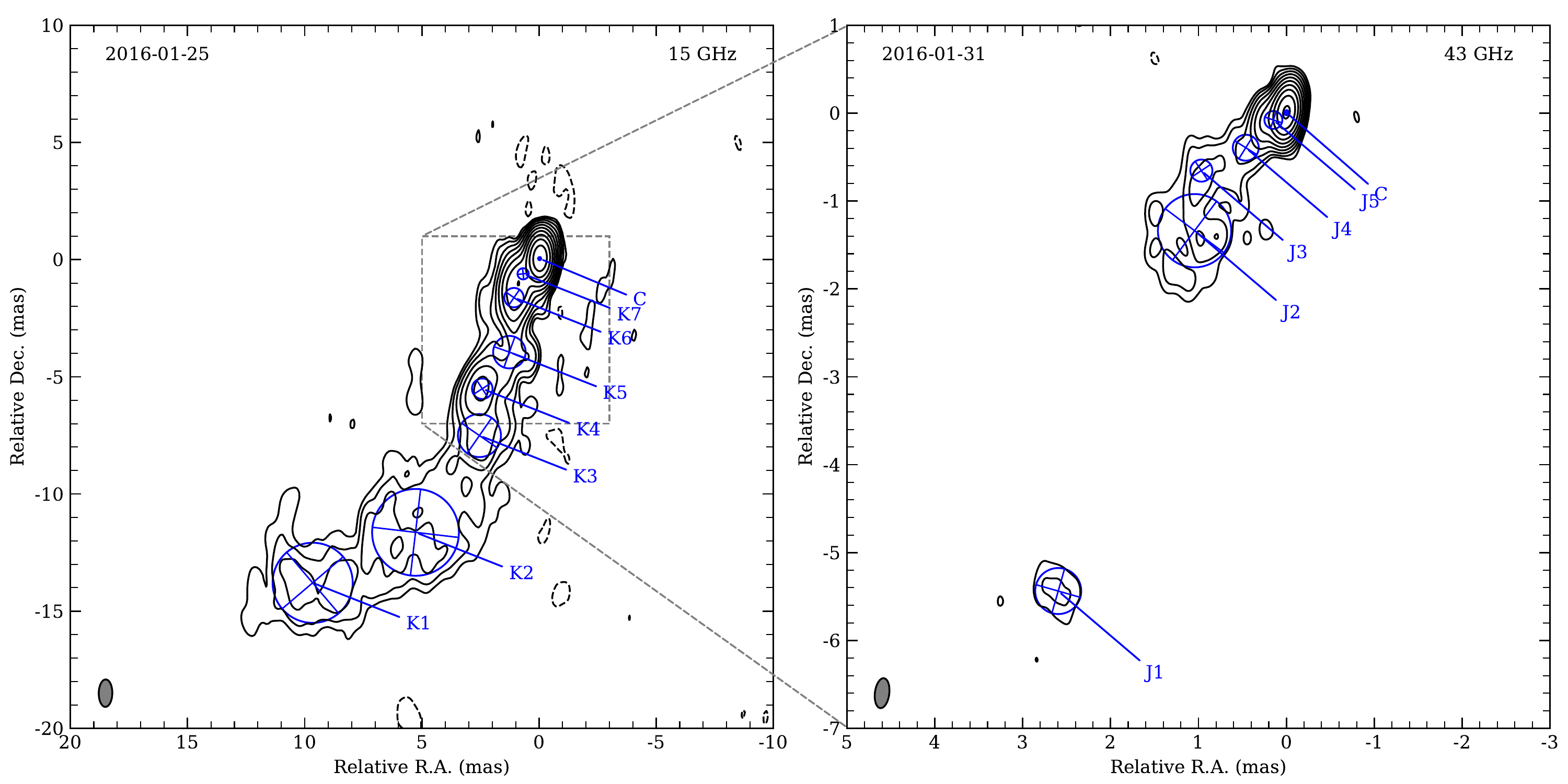}}
\caption{Natural-weighting total intensity image of CTA 102 obtained from 15 GHz VLBA data ({\it left}) and from 43 GHz data ({\it right}).  The 43 GHz observation was made only six days behind the 15 GHz one. The image parameters are referred to Table \ref{tab:obs}. The lowest contour represents 4 times the rms noise ($\sigma_{\rm 15 GHz} = 0.4$ mJy/beam; $\sigma_{\rm 43 GHz} = 0.7$ mJy/beam).  The contours increase in steps of 2. The circles overlaid on the contour map are the fitted Gaussian components. }
\label{fig:15GHz}
\end{figure}

Natural weighting was used for creating the VLBA images. The total intensity image at 15 GHz is presented in Figure \ref{fig:15GHz}-a. The jet extends and expands to the southeast up to a distance of 17 mas (corresponding to a projected length of $\sim$140 pc) and shows multiple twists. The overall jet structure is consistent with the images derived from the Monitoring of Jets in Active Galactic Nuclei with VLBA Experiments program \cite[MOJAVE;][]{2005AJ....130.1389L}, while having a higher sensitivity than the MOJAVE image due to longer integration time and better ({\it u, v}) coverage. The dynamic range (the ratio of peak flux density to image noise) of Figure \ref{fig:15GHz}-a is 3750:1.  A series of bright knots distributed along the ridge line of the jet trace an apparently oscillatory trajectory. Nine jet components are inferred in the 15 GHz image, labelled as K1, K2, ... K9. The model fitting parameters are listed in Table \ref{tab:modelfit}. There is a hint of the jet bending between components K6 and K7 at (R.A., DEC.) $\sim$(1, -1) mas after which the jet gradually expands (more diffuse emission from larger component sizes); another instance of bending occurs between components K2 and K3 at (R.A., DEC.) $\sim $ (4, -9) mas beyond which the expansion becomes more pronounced. Such pc-scale bending and oscillatory features are consistent with either hydrodynamic instability developed on the jet surface as a result of jet-ISM interactions \citep[e.g., 3C~48 jet:][]{1991Natur.352..313W, 2010MNRAS.402...87A} or the jet kinematics being channeled along a helical trajectory \cite[e.g.][]{2013ApJ...768...40L,2016MNRAS.461.3047L}, supported by a strong magnetic field.

The total extent of the 43 GHz jet is 6 mas (a projected size of 50 pc), as is shown in Figure \ref{fig:15GHz}-b, which was derived from the 43 GHz data on 2016 January 31, just six days after the 15 GHz observation. As the other 43-GHz images show similar structure, they are not shown here. The reader can refer to the VLBA-BU-BLAZAR website\footnote{\href{http://www.bu.edu/blazars/VLBA\_GLAST/cta102.html}{\url{http://www.bu.edu/blazars/VLBA\_GLAST/cta102.html}}} which contains all images. The jet components are marked in Figure \ref{fig:15GHz}-b. They were cross-identified from the 43 GHz images based on their similar flux density, size, and position in images in adjacent epochs. Due to resolution and sensitivity difference, not all jet components have one-to-one correspondence at 43 GHz and 15 GHz. A large jet bending from southeast to south happens in the vicinity of J3 component, in a similar way with the 15 GHz structure. The inner jet bending between components J5 and J4 at (R.A., DEC.) $\sim $ (0--0.4,0--0.2) mas is less pronounced here compared to the outer jet, though hints of the oscillatory features are present including in the transition regions between J5, J4, J3 and J2. In addition, the gradual expansion of the jet on moving between J5 and J1 can be seen, consistent with that from the 15 GHz image.

\subsection{Polarization images}
\begin{figure}
\centerline{\includegraphics[height=8.5cm,width=20cm]{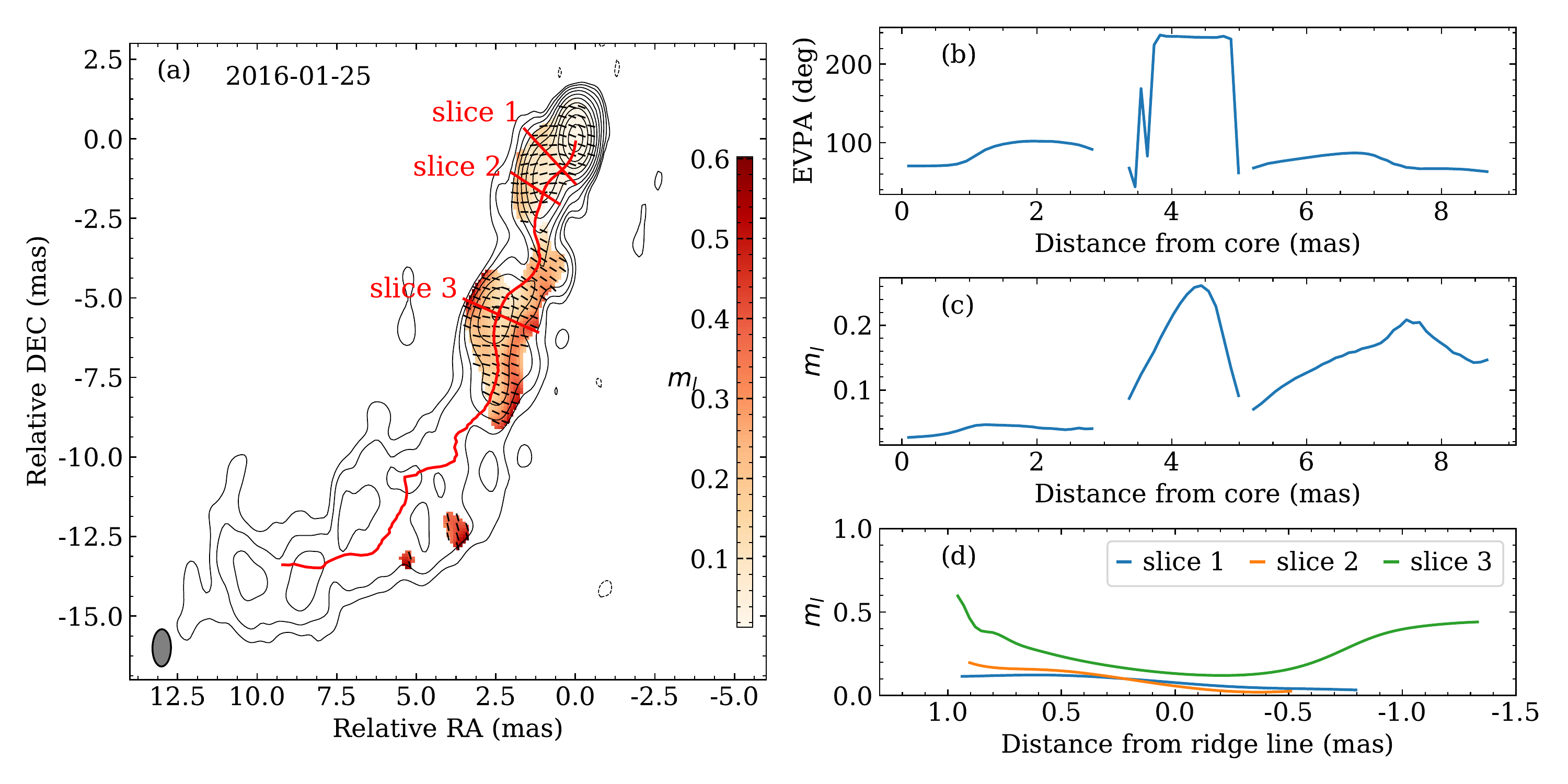}}
\caption{15 GHz polarization image of CTA 102. Panel (a): Linear polarization degrees (pseudo-color image) over the Stokes I contour. The short black bar represent EVPA orientation. The red curve is jet ridge line and the red straight lines are vertical with ridge line. Panel (b): EVPA along jet ridge line. Panel (c): Linear polarization degrees along jet ridge line. (d): Linear polarization degrees along slice 1,2, and 3.}
\label{fig:pol15}
\end{figure}

The 15 GHz polarization image is presented in Figure \ref{fig:pol15}. The uncertainties in polarization flux density and EVPA are $\sim10\%$ and $\sim5\degr$  respectively. It shows a low linear polarization degree ($m_l \sim 4\%$) in the core region and higher $m_l$ in the jet with a maximum of $\sim 60\%$ at (RA, DEC) $\sim$ (3.5, -5) mas. There is also an evidence of enhanced $m_l$ at the outer boundaries. We defined the jet ridge line using the procedure adopted by \citet{2017MNRAS.468.4992P} and investigated the changes of EVPA and $m_{l}$ along the jet ridge line (Figure \ref{fig:pol15}-b). The EVPA at core is $\sim 70\degr$, then monotonously increases to $100\degr$ at a distance of 2 mas, then decreases until $90\degr$ at 4 mas; after that, the EVPA increases again upto $85\degr$ at 7 mas, and returns $\sim 65\degr$ at $r = 8$ mas. The overall variation shows a sinusoidal pattern. The linear polarization degree shows significant changes along the jet ridge line at the locations of rapid EVPA change (Figure \ref{fig:pol15}-c). The highest $m_{l}$ appears at $r = 4.5$ mas, roughly corresponding to the position of minimum EVPA in Figure \ref{fig:pol15}-b. Figure \ref{fig:pol15}-d shows the $m_{l}$ distribution along the three tangent slices to the ridge line passing through jet components K7, K6 and K4. All show weaker polarization at the center and stronger polarization with increasing distance from the center. This may indicate jet-ISM interactions at the boundary, or signatures of a developing pc-scale jet structuring possibly supported by strong magnetic fields in a helical jet scenario.

\begin{figure}
\centerline{\includegraphics[height=10cm,width=10cm]{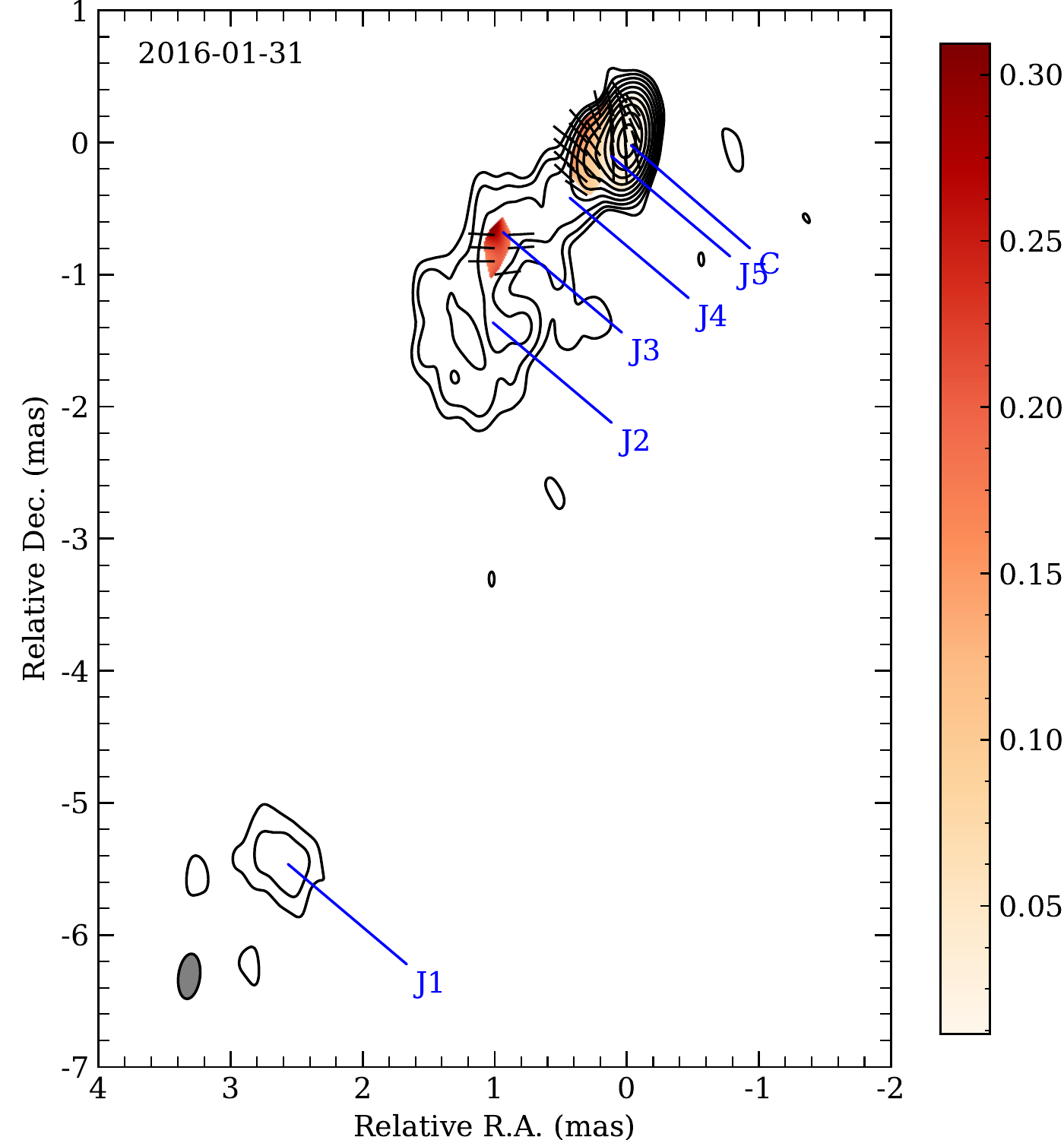}}
\caption{43 GHz polarization image of CTA 102. The linear polarization degrees (pseudo-color image) over the Stokes I contour. The short black bar represent EVPA orientation.}
\label{fig:43pol}
\end{figure}

\begin{figure}
\centerline{\includegraphics[width=0.5\linewidth]{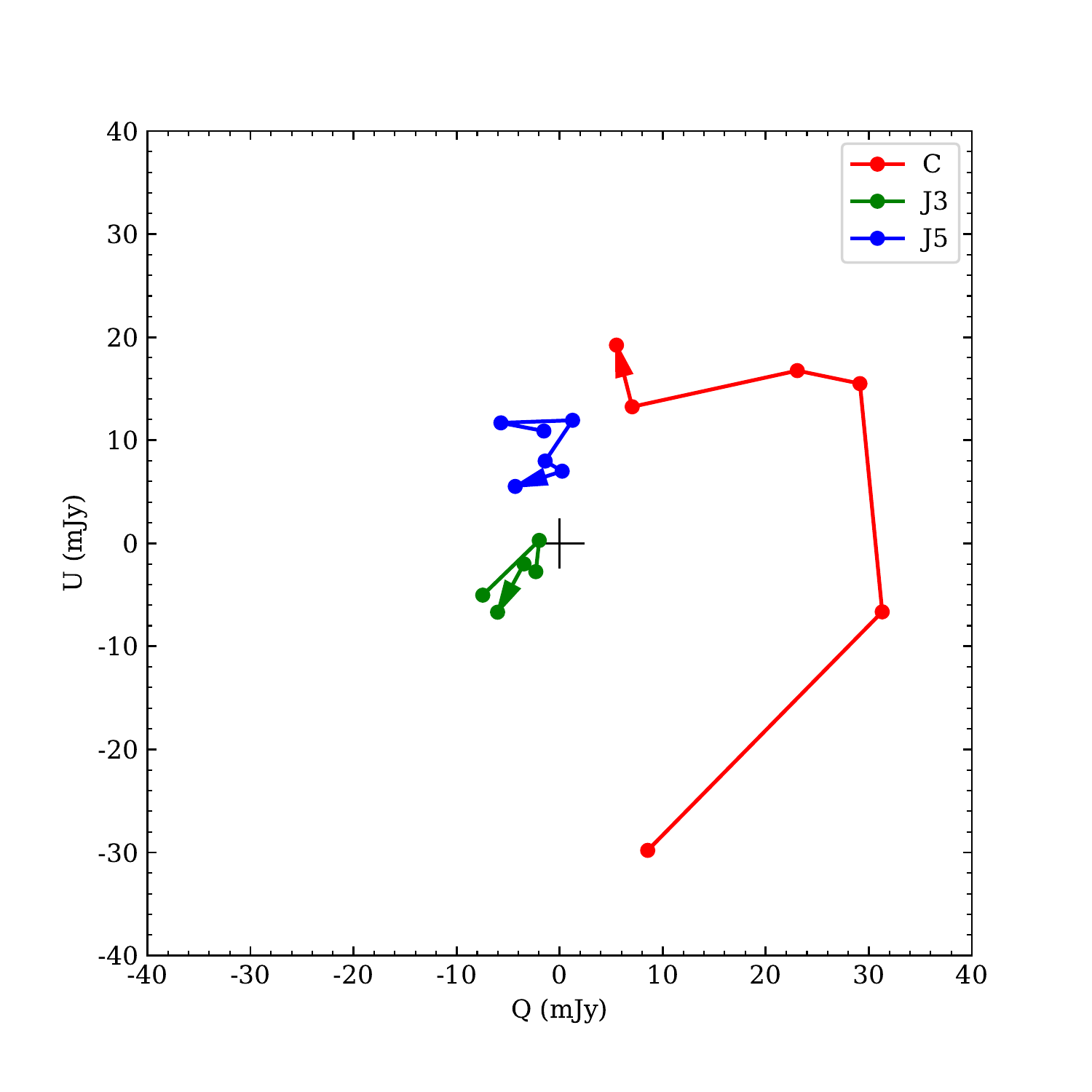}}
\caption{Stokes parameters U vs. Q of different emission components in the jet of CTA 102. The horizontal axis is Stokes Q component and the vertical axis is Stokes U component. The data were plotted here were from December 2015 to June 2016. Component C show a significant rotation, while J3, J5 don't show rapid variation}
\label{fig:pola}
\end{figure}

The 43 GHz polarization image on epoch 2016 January 31 is presented in Figure \ref{fig:43pol}, showing a low polarized core ($m_l \sim 3\%$) consistent with the 15 GHz image. The position of (RA, DEC) $\sim$ (1, -1) mas from the core shows a higher polarization of $20\% - 30\%$. Owing to the weakness of the polarized emission, finer polarization structures tangential or along the jet are not discernible.
Multi-epoch polarization images at 43 GHz from December 2015 to June 2016 show an interesting result that the EVPAs of the core C is rotating anti-clockwise (Figure \ref{fig:pola}). 
 This is consistent with the expectation from a helical jet model in which the Stokes parameters undergo a rotation coincident with a corresponding increase in the polarization degree during a flaring event \cite[e.g.][]{2015ApJ...813...51C}. The EVPAs of jet components J3 and J5 do not show significant changes, but still show a clue of clockwise rotation, in an opposite pattern with the core C. The polarization emission from other jet components are below the image detection threshold.

\subsection{Jet kinematics and geometry}

The apparent transverse velocity $\beta_{\perp}$ (in unit of the speed of light $c$) of a jet knot is related to the intrinsic velocity $\beta$ (in unit of $c$) and the inclination angle $i$ between the jet and the observer's line of sight as
\begin{equation}
\label{eq:vapp}
\beta_{\perp} = \frac{\beta \sin i}{1-\beta \cos i}.
\end{equation}

The extrema of $\beta$, Lorentz factor $\Gamma = (1-\beta^2)^{-1/2}$ and $i$ in terms of $\beta_{\perp}$ are
\begin{align}
\beta &\geq \frac{\beta_{\perp}}{\sqrt{\beta^2_{\perp}+1}}\\ \nonumber
\Gamma &\geq \sqrt{\beta^2_{\perp}+1}\\ \nonumber
i &\leq \cos^{-1} \frac{\beta^2_{\perp}-1}{\beta^2_{\perp}+1}.
\end{align}

\begin{figure}
\centerline{\includegraphics[scale=0.7]{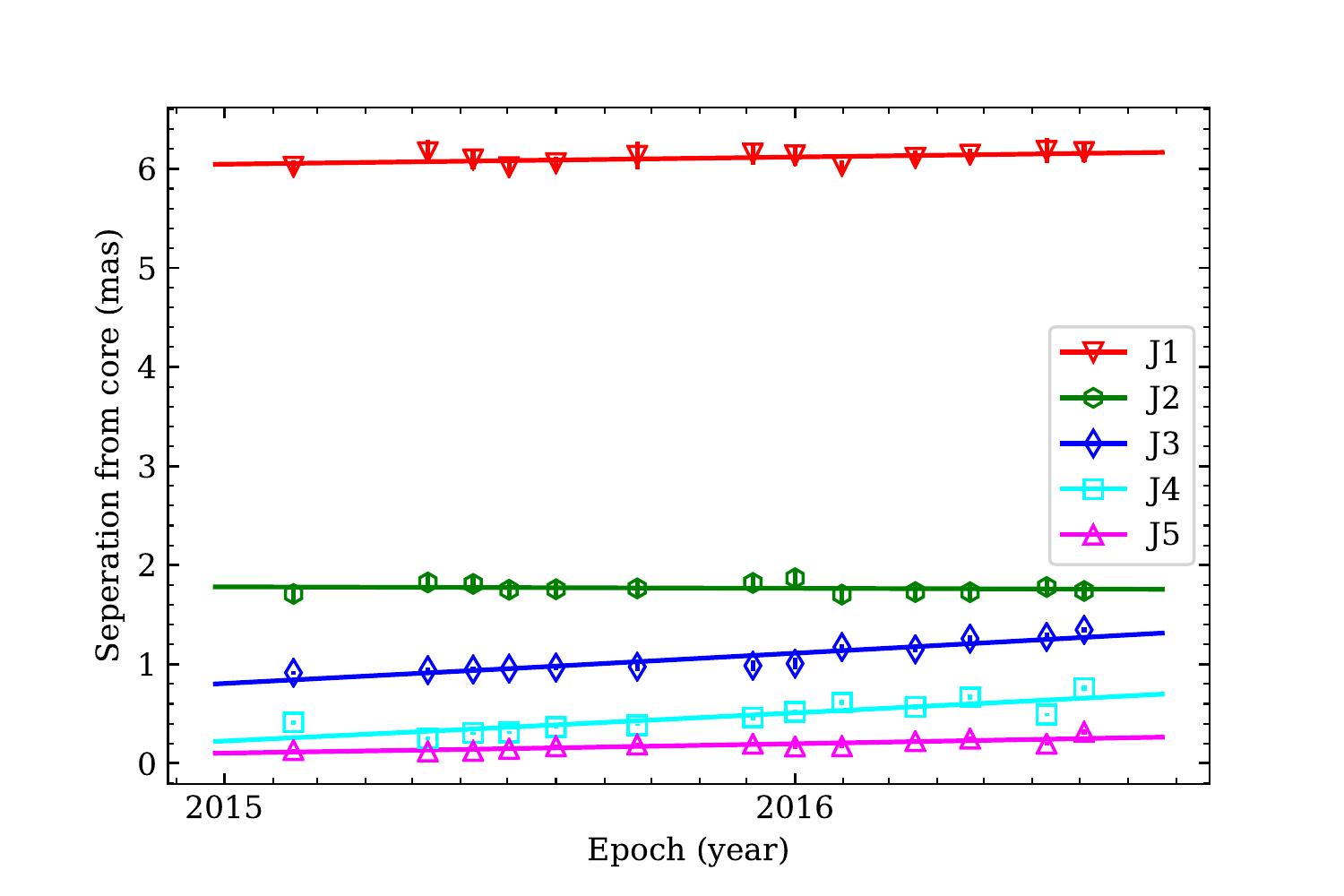}\includegraphics[scale=.6]{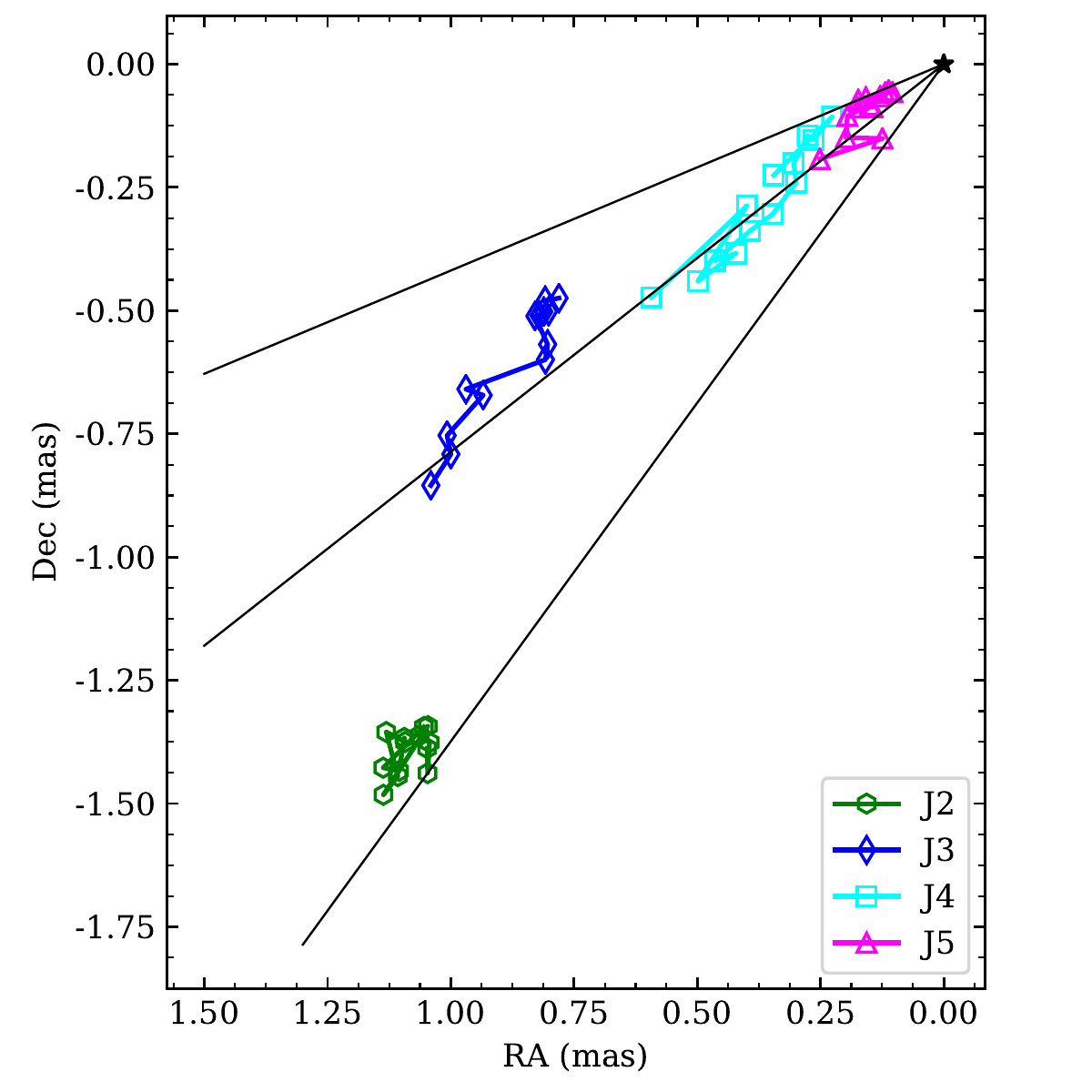}}
\caption{Jet proper motion and half opening angle of CTA 102. Left:  separation of jet component from the core versus observing epoch. The lines represent the linear fit and the slopes correspond to the jet proper motion. Right: Jet half opening angle defined from the inner jet components. The 43-GHz jet positional parameters are adopted.}
\label{fig:ki}
\end{figure}

Figure \ref{fig:ki}-a shows the variation of the jet component separation from the core with time (observation epoch) based on the multi-epoch 43 GHz images. A linear function was used to fit the time dependent separation of components to infer proper motions $\mu$ in the range 0.04 -- 0.33 mas yr$^{-1}$. Component J2 is found on the jet boundary (see Figure \ref{fig:ki}-b); it does not show significant motion during the observations.  The kinematics of J1, inferred from the table and the multi-epoch positions in Figure \ref{fig:ki}-a suggests that it is near stationary, possibly owing to it being a standing shock \citep[e.g., ][]{2009AJ....138.1874L} or at the knot of a helical jet as inferred from the kinematics of the pc-scale components in other similar AGN \cite[e.g.][]{2017ApJ...841..103S,2017ApJ...846...98J}. 

The jet speed indicates a kinematic structuring involving a slower initial jet, an increase and a slowing down such as in an acceleration--deceleration process also seen in other AGN \citep{2005AJ....130.1418J,2016AJ....152...12L}; $\mu {\rm (J5)} = 0.11$ mas yr$^{-1}$ at $\sim$ 0.3 mas increases to $\mu {\rm (J4)} = 0.30$ mas yr$^{-1}$ at $\sim$ 0.66 mas and $\mu {\rm (J3)} = 0.33$ mas yr$^{-1}$ at $\sim$ 1.2 mas; beyond 1.5 mas (corresponding to a de-projected distance of about 106 pc assuming the inclination angle of 6\fdg6), there is a slowing down with $\mu {\rm (J2)} = 0.04$ mas yr$^{-1}$ at $\sim$ 1.8 mas to $\mu {\rm (J1)} = 0.07$ mas yr$^{-1}$ at about 6 mas. The slowest outermost jet component J1 is also the most extended component in 43 GHz data (see Table \ref{tab:modelfit}). From the largest apparent superluminal velocity $\beta_{\perp} = 17.5$, the intrinsic velocity is $\beta \geq 0.998$, bulk Lorentz factor $\Gamma \geq 17.5$ and inclination angle $i \leq 6^{\circ}.6$.

The instantaneous positions (in R.A. and Dec.) of the innermost jet components J3, J4 and J5 upto the jet bending position at $\sim$ 1.5 mas were fitted  with a linear model, shown in Figure \ref{fig:ki}-b. The projected half opening angle is $\psi=15\fdg6$ and the intrinsic half opening angle is therefore $\theta_0 \sim \psi \sin i \leq 1\fdg8$. A mean position angle $\lambda = 128\fdg3$ is determined as the position angle of the fitted central straight line in Figure \ref{fig:ki}-b. These parameters are listed in Table \ref{tab:ki} and used for helical modeling in Section \ref{sec:discussion}. The estimates are similar to the recently updated multi-epoch (1993--2013) study of the pc-scale images and kinematics of the same source by the MOJAVE program \cite[][]{2016AJ....152...12L} which identified six main components spanning between 8 - 37 epochs with $\mu = 0.01 - 0.33$ mas yr$^{-1}$ and $\beta_{\perp} = 0.78 - 17.73$.

\section{Discussion}
\label{sec:discussion}

\subsection{Helical model and optical variability}

The flaring phase studied here (2016-01) leads into a prominent phase towards the end of the year with the helical nature likely becoming more prominent with activity state of the source. CTA 102 underwent an increase of 6 - 7 magnitudes in optical with respect to the general minimum during the end of 2016 and beginning of 2017 when it was monitored using the whole Earth blazar telescope \citep{2017arXiv171202098R}. The study finds that the optical and infra-red emission is composed of both jet based synchrotron (variable) and thermal radiation from the accretion disk (stable, underlying), and that variations in the Doppler factor indicate an origin from orientation changes along the line of sight due to instabilities in the jet resulting in a helical structure. 

A helical jet model \citep[][]{2015ApJ...805...91M} applied previously to quasar jets \citep[PG 1302$-$102; ][]{2016MNRAS.463.1812M} is also employed to model the kinematics of a radiating blob moving on a helical trajectory to simulate variability and the component trajectory for comparison with the observations. The relativistic motion of the jet component is governed by the conservation of the total energy $E$ and the angular momentum $L$ in the jet and a constant half opening angle $\theta_0$ (conical jet). For a jet component launched at a cylindrical distance $\varpi_0$ from the central supermassive black hole with an initial speed $\beta_\perp$ (in units of $c$) and specific angular momentum in units of distance $j = L/(E c)$, the cylindrical distance $\varpi$, vertical height $z$, azimuthal angle $\phi$ and their associated velocities (differential with respect to coordinate time parameter $t$ represented by a dot) in a special relativistic framework \citep{1995A&A...302..335S, 2015ApJ...805...91M} are 
\begin{align}
\varpi &= f \left(1+\left(\frac{a t+b}{f}\right)^2\right)^{\frac{1}{2}};~ \dot{\varpi} = \frac{a}{\varpi} (a t+b), \\ \nonumber
z &= \frac{\varpi - \varpi_0}{\tan \theta_0};~ \dot{z} = \frac{\dot{\varpi}}{\tan \theta_0},\\ \nonumber
\phi &= \frac{1}{\sin \theta_0} \left ( \tan^{-1} \frac{a t+b}{f}-\tan^{-1} \frac{b}{f} \right);~\dot{\phi} = \frac{a f}{\varpi^2 \sin \theta_0},
\end{align}
where $a = \beta \sin \theta_0$, $b = (\varpi^2_0-j^2/\beta^2)^{1/2}$, $f = j/\beta$ and the coordinate time $t$ expressed in units of distance. The angle between the observer's line of sight and the direction of the instantaneous velocity vector of the jet component $\xi$ is given by
\begin{equation}
\cos \xi = \frac{\dot{\varpi} \cos \phi \sin i-\varpi \dot{\phi} \sin \phi \sin i+\dot{z} \cos i}{(\dot{\varpi}^2+\varpi^2 \dot{\phi}^2+\dot{z}^2)^{1/2}},
\end{equation}
that modulates the Doppler factor
\begin{equation}
D = \frac{1}{\Gamma (1-\beta \cos \xi)},
\end{equation}
where $\Gamma$ is the jet bulk Lorentz factor for a bulk velocity $\beta$. The relativistically beamed flux density from a resolved jet component $S_{\nu,0}$ in the instantaneous rest frame is received by a distant observer as
\begin{equation}
S_\nu = D^{3-\alpha} S_{\nu,0},
\label{flux}
\end{equation}
where $\alpha$ is the spectral index in the observer's frame. For synchrotron emission from an optically thin jet, the polarization degree in the case of a helical magnetic field geometry \citep{2005MNRAS.360..869L,2013MNRAS.436.1530R} is
\begin{equation}
m_{l} \sim m_{l, \rm max} \sin^2 \xi',
\end{equation}
where $\xi'$ is the viewing angle in the jet frame and is related to $\xi$ by
\begin{equation}
\sin \xi' = \frac{\sin \xi}{\Gamma (1-\beta \cos \xi)}.
\end{equation}
The instantaneous pitch angle between the spiral motion and the helical field is $\tan \zeta = \dot{z}/(\varpi \dot{\phi})$ using which, the EVPA is
\begin{equation}
\tan \chi = \frac{\zeta \sin \phi}{\zeta \cos \phi-i},
\end{equation}
and the Stokes parameters $Q$ and $U$ are expressed in terms of the polarization degree and the EVPA as
\begin{equation}
Q = m_{l} \cos 2 \chi; \ U = m_{l} \sin 2 \chi.
\end{equation}
The instantaneous jet component position (in the coordinate frame) is transformed to the observer's Cartesian coordinates using Euler rotation matrices. In terms of the inclination angle $i$ and jet position angle $\lambda$, the position $(x,y)$ is
\begin{align}\label{xy}
x &= \varpi \cos \phi \cos i \cos \lambda+z \sin i \cos \lambda-\varpi \sin \phi \sin \lambda,\\ \nonumber
y &= \varpi \cos \phi \cos i \sin \lambda+z \sin i \sin \lambda-\varpi \sin \phi \cos \lambda.
\end{align}
The time interval $d \tau$ in a local observer frame is related to the coordinate time interval $dt$ by
\begin{equation}
\tau = (1+\tilde{z}) \int^{t}_{0} (1-\beta \cos \xi) dt,
\end{equation}
accounting for the relative motion between the moving jet component and the distant observer and for cosmological expansion through the redshift factor $\tilde{z}$. The instantaneous flux density variation $S_\nu$, polarization degree $m_{l}$, EVPA $\Theta$, and position $(x,y)$ are determined in the coordinate frame and re-cast into the observer frame for comparison with observed data.

For the simulations, kinematic parameters adopted are $\beta_0 = 0.998$, $\lambda = 128\fdg3$, the equalities in the limits $\Gamma = 17.5$, $i = 6\fdg6$, and $\theta_{0} = 1\fdg8$; 
 emission parameters include a multi-band optical (BVRI)--near infra-red (JHK) spectral index $\alpha \sim 1.5$ inferred during the strong 2012 September - October outburst \citep{2016MNRAS.461.3047L}, and a maximum optical polarization degree $\Pi_{\rm max} = 0.3$; the component launch radius is constrained based on the minimum separation distance from the core such that $\varpi_0 \leq 0.53$ pc; and, the specific angular momentum $j$ is limited by $j \leq \beta \varpi$ in the simulation to obtain real valued minima.

With an observed relative separation from the core $\omega = (R.A.^2+DEC.^2)^{1/2}$ and helical jet model based expectation $\omega_M = (x^2+y^2)^{1/2}$ from equation (\ref{xy}), the residual distance is
\begin{equation}
D_\omega = |\omega-\omega_M|.
\end{equation}

\begin{figure}
\includegraphics[scale=0.21]{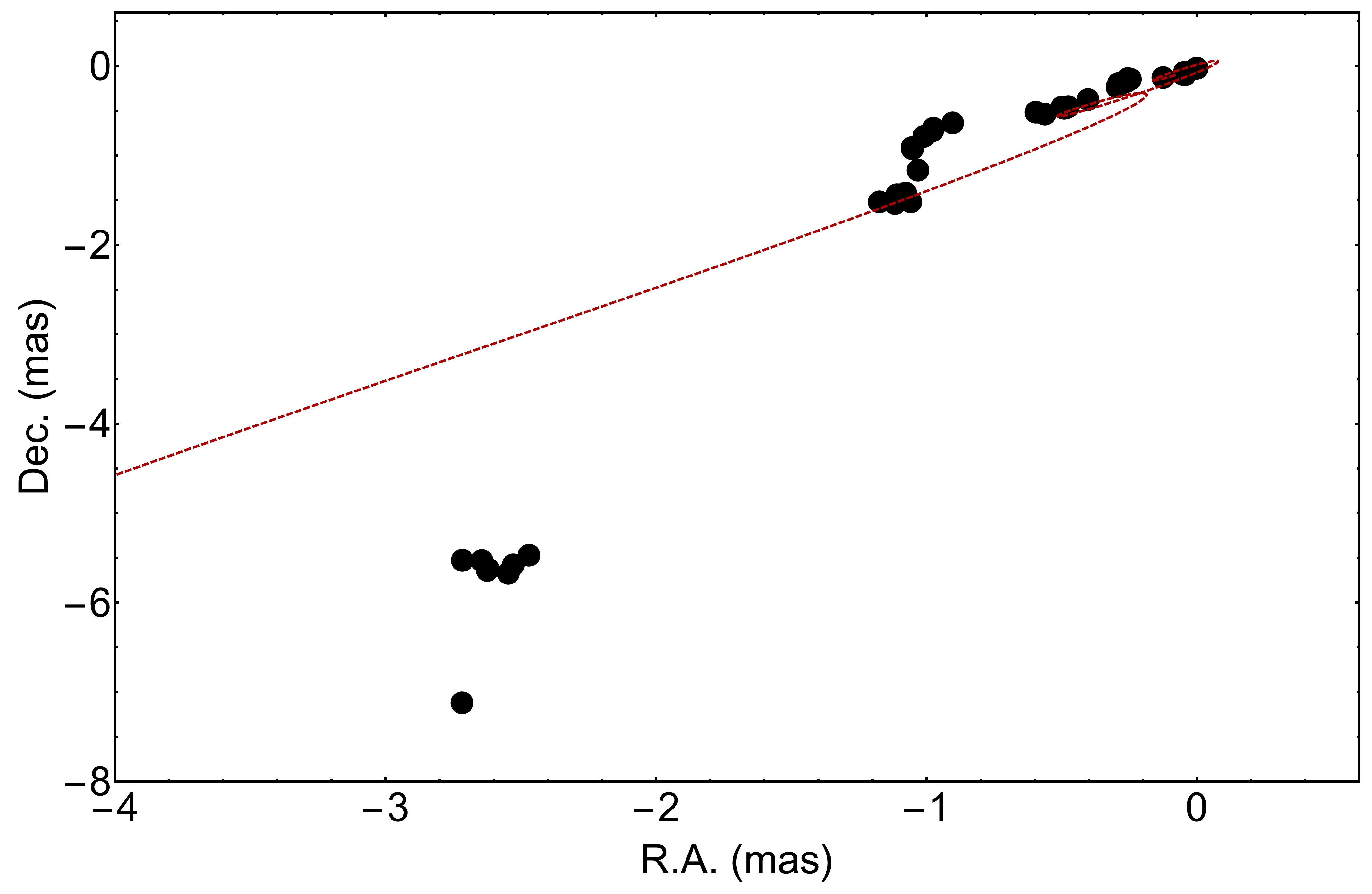}\includegraphics[scale=0.21]{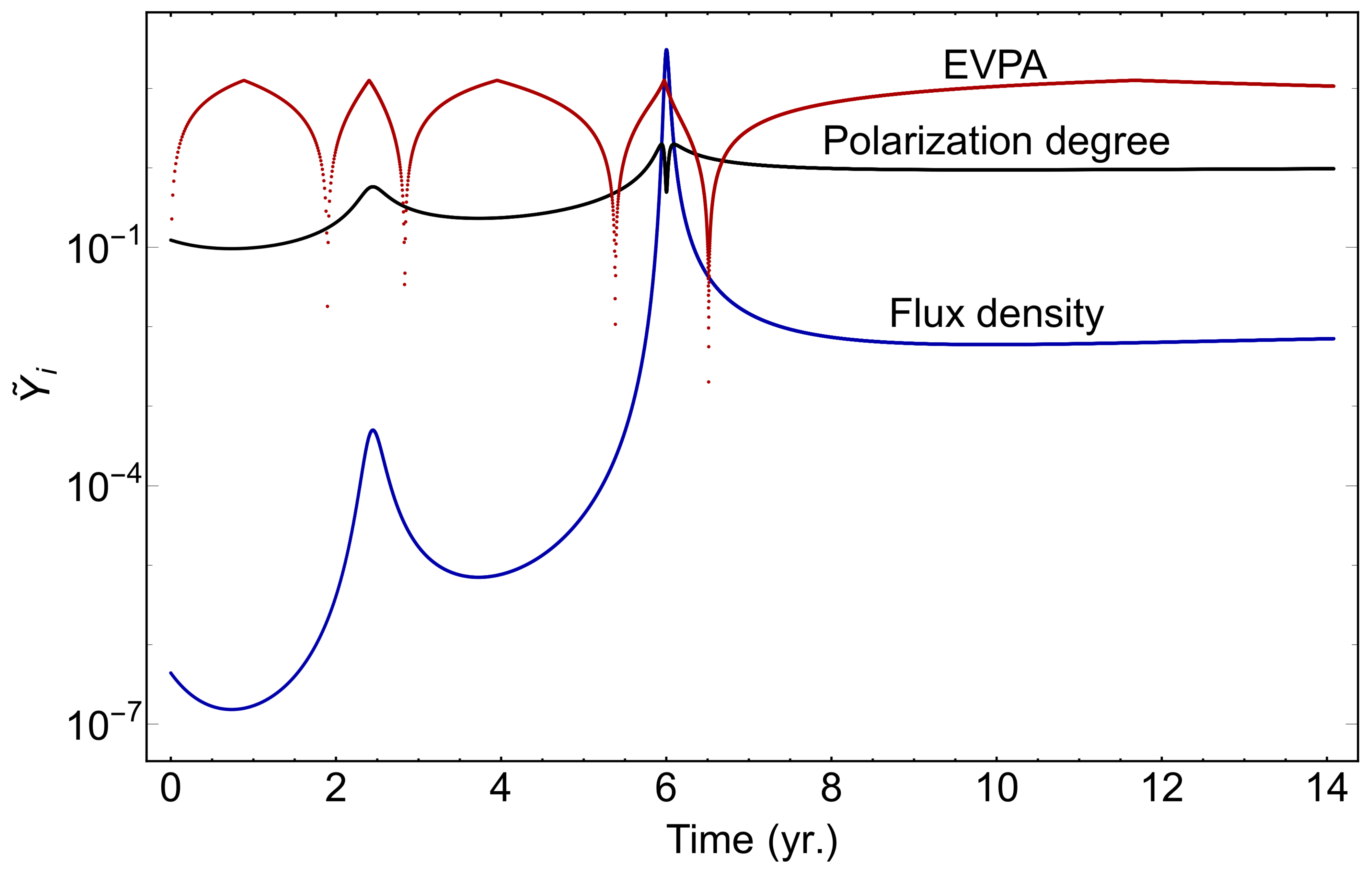}\\
\centering\includegraphics[scale=0.22]{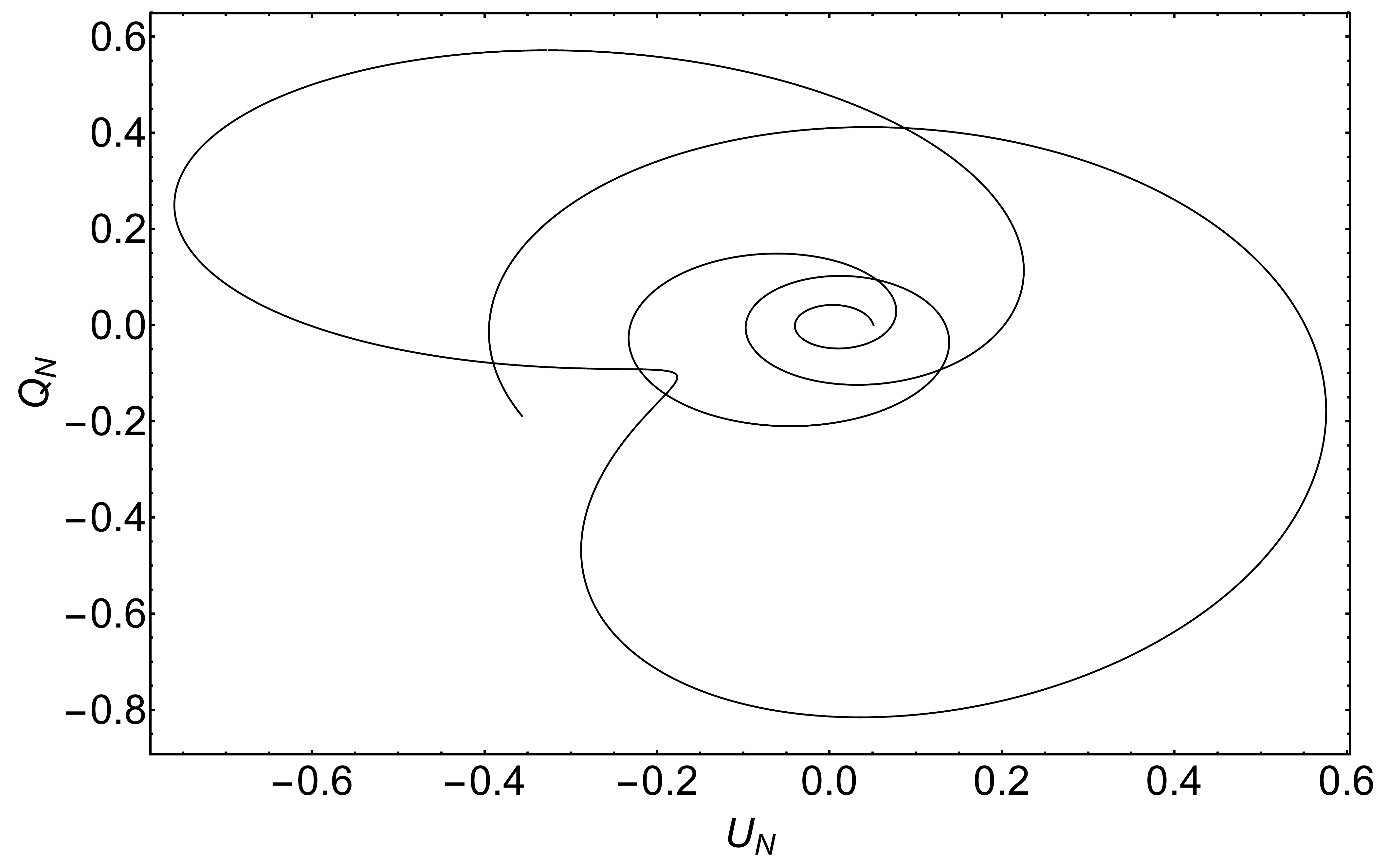}
\caption{Simulation of CTA 102 jet component trajectory, flux density, polarization degrees, EVPA, and Stokes parameters. Top left: Comparison of the simulated the component trajectory with observed values (black colored dots). Top right: The simulated normalized flux density (blue), polarization position angle (black), EVPA (red). Bottom: The rotation of the Stokes Q and U parameters. A comparison of the contrast ratio of the simulations and from the data suggests that flux density and polarisation degree variability are well described by the long term helical jet activity while EVPA variability is better described by short term mechanisms.}
\label{fig:XYplots}
\end{figure}

This is subjected to a constrained minimization at each observation epoch $t$. Mean values of $\varpi_0$ = 0.53 pc and $j$ = 0.25 pc are obtained and used as inputs to simulate the the component trajectory (Figure \ref{fig:XYplots}-a), variable flux density, linear polarization degree and EVPA in the core-jet (Figure \ref{fig:XYplots}-b), and the rotation of the polarisation vector in the Stokes Q-U plane (Figure \ref{fig:XYplots}-c). As the fit is being done with the jet component positions, the expected trends in the helical model are applicable to long timescale (years) variability. The simulated component trajectory indicates a qualitative agreement with the observed multi-epoch positions of the innermost components J2 - J5 with (R.A.,Dec.) $\leq$ (1.5, 2.0) mas. The outer component J1 at (R.A., Dec.) $\sim$ (1.5, 2.0) mas is however not contained within the helical envelope suggesting that J1 is a feature along the jet wall or moving towards us, and indicates slow kinematics due to the projection effects with a very small motion perpendicular to our line of sight (proper motion).

 The comparative properties are shown in Fig. \ref{fig:XYplots} and are cast in terms of the quantities $\tilde{Y}_i = Y_{i}/{\rm Mean}(Y_{i})$ where the subscript $i$ quantifies the contribution of helical motion based variability in flux density, polarization degree and EVPA respectively. Four important implications can be drawn for comparison with observations. The first is the contrast between the maximum and minimum of $\tilde{Y}_i$ in units of its standard deviation $C_{\rm sim.} = ({\rm Max}(\tilde{Y}_i)-{\rm Min}(\tilde{Y}_i))/\sigma_{Y_i}$ = 
 (7.0,5.1,2.8) over the full simulation duration ($\sim$ 14 yr.); the corresponding contrasts from the observations $C_{\rm obs.} = ({\rm Max}(\tilde{X}_i)-{\rm Min}(\tilde{X}_i))/\sigma_{X_i}$ = (4.2, 4.2, 4.5). The results can be used to comment on the relative influences of physical processes causing the long term (year timescale; helical jet scenario) and short term (months) variability; that in flux density and polarisation degree over the long term (year timescale) is sufficiently captured by the helical scenario; and that in EVPA is dominated by short term mechanisms. The second is the typical timescale between prominent flares in the model which is $\sim$ 2.5 - 3.5 yr suggesting again that the helical signatures are operational over long timescales. The third relates to the qualitative relation between the phases of the variability in flux density, polarisation degree and EVPA; one finds correlated, anti-correlated variability and intermediate phase shifts, consistent with the observations. The fourth is in terms of the expected anti-clockwise rotation of the Stoke's parameters Q and U in the helical jet scenario including an increasing amplitude with increasing variability cycle; a similar pattern with minor departures (possibly due to turbulence or other short timescale variability processes) is inferred in the observed changes of Q and U.

 Over short timescales such as the duration spanned in the current study, additional variability mechanisms may be operational including magnetic reconnection events causing flares \cite[e.g.][]{2003NewAR..47..513L} or shocks in jet \cite[e.g.][]{1985ApJ...298..114M} such as during injection events with the beamed emission at suitable viewing angles. It is then required for systematic studies over more observational epochs (simultaneous observations of flux density, polarization and high-resolution imaging are desirable) and spanning a long duration for a clearer picture to emerge. Anti-correlated variability between flux density and polarization degree has been reported \cite[e.g.][]{2014ApJ...781L...4G} indicating a consistency of the simulations with the observations in this study.

\subsection{Location of the $\gamma$-ray flare}
\label{sec4.2}

Assuming a typical $\gamma$-ray flare timescale $\Delta t$ of 10 days - 1 month (Figure \ref{fig:lc}-top), and using a Doppler factor $\delta \sim \Gamma = 17.5$ and redshift $\tilde{z} = 1.037$, the size of the compact region participating in the variability $\Delta r \leq c \delta \Delta t/(1+\tilde{z}) = 0.11 - 0.32$ pc. For a conically shaped relativistic jet, the jet opening angle is $\tan \theta_0 \sim \Delta r/r$ from which the distance of this region from the central engine is $r = 5.7 - 16.7$ pc. Unlike in the case of young and un-beamed AGN where it was argued that MeV $\gamma$-ray emission can arise from jet-ISM interaction in the kpc-scale lobes \citep{2017MNRAS.466..952A}, for regular radio-loud beamed AGN, MeV--GeV band $\gamma$-ray emission can arise from shocked regions in the jet due to injection events just outside the broad line region and upto pc scales \cite[e.g.][]{2017MNRAS.468.4478L} with a large contribution from the Compton up-scattering of lower energy photons \cite[e.g.][]{2015ApJ...813...51C,2017arXiv170909342G}. These events can produce short timescale variability (days to months in observer's frame) as observed in the $\gamma$-rays and optical R-band. The long-term year-scale variability may then be attributable to the helical jet, consistent with the indications for this source \cite[e.g.][]{2013A&A...557A.105F,2015ApJ...813...51C,2016MNRAS.461.3047L}.

\subsection{Core shift effect}

The new 15 GHz observation on 2016 January 25 preceeds the 43 GHz observation by six days. The availability of quasi-simultaneous 15 and 43 GHz observations covering the $\gamma$-ray flaring period can be used to estimate the pc-scale core distance and magnetic field strength in the context of the core-shift effect \cite[e.g.][]{1998A&A...330...79L, 2005ApJ...619...73H, 2015MNRAS.452.2004M,2017MNRAS.469..813A}. The estimated magnetic field strength can provide clues on the typical conditions influencing the pc-scale helical jet motion. The observed core-shift is an apparent shift in the position of the emitting core ascribed to optical depth based effects. Synchrotron self-absorption in the core results in a steep power-law decline of the radio spectrum following a rise and turnover \cite[e.g.][]{2013A&A...557A.105F}. Assuming a conical jet geometry and equipartition between the magnetic energy density and the particle kinetic energy density in the pc-scale jet, the core offset per unit observation frequency $\Omega_{r \nu}$ (pc GHz), core distance $r_{\rm core}$ (pc) and the magnetic field strengths at 1 pc ($B_1$ in G) and at the core ($B_{\rm core}$ in G) are
\begin{align}
\Delta \theta &= \Delta d \sin i / \tan \theta_0\\
\Omega_{r \nu} &= 4.85\times10^{-9} \frac{D_L \Delta \theta}{(1+\tilde{z})^2 \left(\nu^{-1}-\nu^{-1}_0\right)},\\
r_{\rm core} &= \frac{\Omega_{r \nu}}{\nu \sin i},\\
B_{1} &\cong 0.025 \left(\frac{\Omega_{r\nu}^3 (1+\tilde{z})^2}{\Gamma^2 \theta_0 \sin^2 i}\right)^{1/4},\\
B_{\rm core} &= B_{1} \left(r_{\rm core}/(1~{\rm pc})\right)^{-1}.
\end{align}
where $D_L$ is the luminosity distance, $\tilde{z}$ is the redshift, $\nu_0$ is a reference observation frequency, and $\Delta \theta$ is the difference between the apparent core position measured at frequencies $\nu$ and $\nu_0$. From Table \ref{tab:modelfit}, $d_{\rm core,15 GHz} = 0.14$ mas for epoch 2016 January 25 and $d_{\rm core,43 GHz} = 0.06$ mas from which $\Delta d \sim 0.08$ mas. Using $D_L = 6942.2$ Mpc, $\tilde{z} = 1.037$, $\nu = 15$ GHz, $\nu_0 = 43$ GHz; and Doppler factor $D = \Gamma = 17.5$, $\theta_0 = 1\fdg8$ and $i = 6\fdg6$ from Table \ref{tab:ki}, we derived $\Omega_{r \nu} = 40.5$ pc GHz, $r_{\rm core, 43GHz} = 22.9$ pc, $B_1 = 0.96$ G and $B_{\rm core, 43GHz} = 0.04$ G, which is roughly consistent with the estimated $B_{\rm core} = 0.07 - 0.11$ G in \citet{2013A&A...557A.105F} for this source. The study of \cite{2012A&A...545A.113P} employed 8 and 15 GHz observations to estimate a larger $\Delta \theta = 0.32$ mas resulting in estimates of $\Omega_{r \nu} = 46.48$ pc GHz, $r_{\rm core, 15 GHz} = 46.7$ pc, $B_1 = 2.12$ G and $B_{\rm core, 15 GHz} = 0.05$ G, in agreement with expectations from the synchrotron self absorption scenario. The study of \cite{2016ApJ...826..135L} used core brightness temperature estimates at 2, 8, 15 and 86 GHz and synchrotron luminosity to infer jet kinematic parameters and hence $B_1 = 4.64$ G which is a factor of $\sim$5 times our estimate, possibly due to an under-estimation of the jet kinematic parameters. The derived $r_{\rm core}$ is larger than the estimate of the distance of the $\gamma$-ray emitting zone ($r = 5.7 - 16.7$ pc in Section \ref{sec4.2}), implying that the 2016 January $\gamma$-ray flare arises from an inner jet upstream the core and the current VLBA 43 GHz is still not able to resolve that compact region.

\section{Summary} \label{sec:summary}

The presented 15 GHz VLBA observations were carried out on 2016 January 25, during a prominent $\gamma$-ray flare with quasi-simultaneous monitoring also at 43 GHz in an ongoing survey of $\gamma$-ray blazars by the Boston University group.
The main results from this study include:
\begin{enumerate}

\item An oscillatory and bending pc-scale ($\leq$ 17 mas) jet structure is inferred from the 15 and 43 GHz multi-epoch VLBA images spanning $\sim$ 17 months. 

\item Proper motions for the innermost ($\leq$ 1 mas) jet components (J3, J4, J5) were determined in the range of 0.04 -- 0.33 mas yr$^{-1}$. The jet proper motions were employed to estimate the maximum bulk Lorentz factor $\Gamma \geq 17.5$, mean jet position angle $\lambda = 128\fdg3$, inclination angle $i \leq 6\fdg6$ and intrinsic half opening angle $\theta_0 \leq 1\fdg8$.

\item The 15 and 43 GHz polarization images indicate a weakly polarized core and moderately polarized jet components. The polarization is observed to increase along the jet walls, likely manifesting the helical magnetic field.

\item A helical jet model was applied to simulate long-term optical-band variability. 
The contrast in estimates for flux density, polarization degree and EVPA from the simulation suggest that long term variability is sufficiently captured in the helical scenario. 
A suggested typical duration between two strong flares is $\sim$ 2.5 - 3.5 years.
There are patterns of correlated and anti-correlated variability between the polarization degree and the flux density, in agreement with the current observations as well as with other studies. 
A developing observed anti-clockwise rotation of the polarization vector in the Stokes Q-U plane is consistent with expectation from the simulations.

\item An oscillatory pc-scale jet morphology, polarization behaviour and the expectation of $\gamma$-ray emission from the pc-scales are employed to argue for a long timescale (years) dominance by the helical jet scenario with kinematics being supported by a magnetic surface.

\item Quasi-simultaneous observations at 15 GHz and 43 GHz are employed to estimate an apparent core shift of $\Omega_{r \nu} = 40.5$ pc GHz using which the distance of the radio core from the jet base $r_{\rm core, 43GHz} = 22.9$ pc, the magnetic field strength at the pc-scale $B_1 = 0.96$ G and at the core $B_{\rm core, 43GHz} = 0.04$ G are estimated. The derived magnetic field strength well agrees with the estimated $B_{\rm core} = 0.07 - 0.11$ G \citep{2013A&A...557A.105F} for this source.

\end{enumerate}

The utility lies in the estimation of physical parameters of the pc-scale jet and the ability to identify and distinguish between mechanisms resulting in the observational signatures. Such high resolution observations reveal a complex kinematic structure which can offer clues of the launching, acceleration and stability of the nascent relativistic jet and its relation to the central engine.

\section*{Acknowledgements}
We thank the referee for suggesting modifications which helped improving our manuscript. TA thanks the grant supported by the Youth Innovation Promotion Association of CAS and FAST Fellowship by the Center for Astronomical Mega-Science, CAS. PM is supported by the CAS-PIFI (grant no. 2016PM024) post-doctoral fellowship and the NSFC Research Fund for International Young Scientists (grant no. 11650110438). ZLZ thanks the Hundred Talents Program of the CAS. We thank Ken Kellerman for helpful comments on the manuscript. TA thanks Svetlana Jorstad for providing the calibrated 43-GHz VLBA data. PM thanks Arun Mangalam for useful discussions on polarisation variability in the helical jet scenario. XFL thanks Sandor Frey for helpful comments on the manuscript and Mai Liao for her help on optical data reduction. PM and TA thank Willem Baan for useful discussions. The VLBA experiment resulting in the 15 GHz image is sponsored by Shanghai Astronomical Observatory through the MoU with the NRAO. This study makes use of 43 GHz VLBA data from the VLBA-BU Blazar Monitoring Program (VLBA-BU-BLAZAR; \href{http://www.bu.edu/blazars/VLBAproject.html}{http://www.bu.edu/blazars/VLBAproject.html}), funded by NASA through the Fermi Guest Investigator Program. During the observations reported here, the VLBA was an instrument of the National Radio Astronomy Observatory. The National Radio Astronomy Observatory is a facility of the National Science Foundation operated by Associated Universities, Inc. This research has made use of data from the MOJAVE database that is maintained by the MOJAVE team (Lister et al., 2009, AJ, 137, 3718). Data from the Steward Observatory spectro-polarimetric monitoring project were used. This program is supported by Fermi Guest Investigator grants NNX08AW56G, NNX09AU10G, NNX12AO93G, and NNX15AU81G.
\software{AIPS \citep{1996ASPC..101...37V}}\\
\software{Difmap \citep{1997ASPC..125...77S}}.

\begin{table}
\centering
\caption{VLBA observation logs. Columns are as follows: (1) date of observation, (2) experiment code, (3) observing frequency, (4) bandwidth, (5) total on source time, (6) total flux density, (7) peak flux, (8) rms noise level of image, (9 - 10) major and minor axes of the restoring beam, (11) position angle of the major axis, measured from the north to the east.}\label{tab:obs}
\begin{tabular}{llllllllllll} 
\hline \hline
Epoch & Code & $\nu$ & BW & $t_{\rm int}$ & $S_{rm tot}$ & $S_{\rm peak}$ & rms & $b_{maj}$ & $b_{min}$ & P.A.\\
(yyyy-mm-dd) & & (GHz) & (MHz) & (min.) & (Jy) & (Jy/beam) & (mJy/beam) & (mas) & (mas) & ($\deg$) \\
(1) & (2) & (3) & (4) & (5) & (6) & (7) & (8) & (9) & (10) & (11)  \\ \hline
2015-02-14 & BM413E & 43 & 256 & 45 & 1.9 & 1.5 & 0.9 & 0.69 & 0.31 & 21.1 \\
2015-05-11 & BM413G & 43 & 256 & 50 & 2.3 & 1.6 & 0.7 & 0.54 & 0.24 & 16.5 \\
2015-06-09 & BM413H & 43 & 256 & 42 & 2.2 & 1.5 & 0.7 & 0.38 & 0.16 & -7.7 \\
2015-08-01 & BM413J & 43 & 256 & 40 & 2.1 & 1.5 & 0.7 & 0.41 & 0.20 & -8.6 \\
2015-09-22 & BM413K & 43 & 256 & 42 & 2.3 & 1.7 & 0.8 & 0.40 & 0.18 & -7.0 \\
2015-12-05 & BM413L & 43 & 256 & 44 & 2.2 & 1.8 & 0.6 & 0.54 & 0.22 & -9.5 \\
2016-01-01 & BM413M & 43 & 256 & 46 & 2.6 & 2.0 & 0.7 & 0.55 & 0.18 & $-$12.2 \\
2016-01-25 & BA113C & 15 & 256 &360 & 2.7 & 1.5 & 0.4 & 1.17 & 0.58 & $-$0.8 \\
2016-01-31 & BM413N & 43 & 256 & 43 & 2.3 & 1.6 & 1.1 & 0.34 & 0.16 & $-$7.1 \\
2016-03-18 & BM413O & 43 & 256 & 45 & 2.3 & 1.8 & 0.7 & 0.44 & 0.18 & $-$5.9 \\
2016-04-22 & BM413P & 43 & 256 & 44 & 2.2 & 1.8 & 0.7 & 0.37 & 0.16 & $-$6.8 \\
2016-06-10 & BM413Q & 43 & 256 & 49 & 3.0 & 2.4 & 0.7 & 0.42 & 0.17 & $-$4.7 \\
2016-07-04 & BM413R & 43 & 256 & 44 & 2.9 & 2.5 & 0.8 & 0.45 & 0.19 & $-$10.7 \\ \hline
\end{tabular} \\
Note: on epoch 2016 January 25, HN station did not participate in the observation.
\end{table}



\begin{center}
\begin{longtable}{ccccccc}
\caption{Model fitting results. (1) observation epoch, (2) observing frequency in GHz, (3) jet component label, (4) flux density, (5) position offset from the core component, (6) position angle with respect to the core component, (7) full width of half maximum (FWHM) of the fitted circular Gaussian model.}\label{tab:modelfit}\\ \hline \hline
Epoch & $\nu$ & Comp. & Flux density & r   & PA  & d \\ 
(yyyy-mm-dd) & (GHz)   &  & (mJy)  & (mas) & (deg.) & (mas) \\
(1) & (2) & (3) & (4) & (5) & (6) & (7) \\
\hline
2015-02-14 & 43 & C & 1281.6$\pm$64.1 & 0 & 0 & 0.065$\pm$0.003 \\
2015-02-14 & 43 & J5 & 362.1$\pm$18.1 & 0.132$\pm$0.008 & 116.1$\pm$3.5 & 0.077$\pm$0.004 \\
2015-02-14 & 43 & J4 & 64.6$\pm$3.2 & 0.412$\pm$0.025 & 123.2$\pm$3.6 & 0.253$\pm$0.013 \\
2015-02-14 & 43 & J3 & 72.8$\pm$3.6 & 0.914$\pm$0.015 & 121.3$\pm$1.0 & 0.152$\pm$0.008 \\
2015-02-14 & 43 & J2 & 133.6$\pm$6.7 & 1.709$\pm$0.074 & 141.9$\pm$2.5 & 0.745$\pm$0.037 \\
2015-02-14 & 43 & J1 & 41.8$\pm$2.1 & 6.021$\pm$0.067 & 154.8$\pm$0.6 & 0.671$\pm$0.034 \\ \hline
2015-05-11 & 43 & C & 1343.8$\pm$67.2 & 0 & 0 & 0.033$\pm$0.002 \\
2015-05-11 & 43 & J5 & 450.1$\pm$22.5 & 0.119$\pm$0.005 & 118.6$\pm$3.4 & 0.052$\pm$0.003 \\
2015-05-11 & 43 & J4 & 215.0$\pm$10.7 & 0.250$\pm$0.017 & 115.3$\pm$4.5 & 0.174$\pm$0.009 \\
2015-05-11 & 43 & J3 & 85.3$\pm$4.3 & 0.940$\pm$0.026 & 120.7$\pm$1.6 & 0.259$\pm$0.013 \\
2015-05-11 & 43 & J2 & 138.5$\pm$6.9 & 1.825$\pm$0.082 & 141.5$\pm$2.6 & 0.819$\pm$0.041 \\
2015-05-11 & 43 & J1 & 75.6$\pm$3.8 & 6.168$\pm$0.121 & 155.2$\pm$1.1 & 1.208$\pm$0.060 \\ \hline
2015-06-09 & 43 & C & 1270.9$\pm$63.5 &0 & 0 & 0.024$\pm$0.001 \\
2015-06-09 & 43 & J5 & 578.1$\pm$28.9 & 0.124$\pm$0.011 & 115.5$\pm$5.4 & 0.113$\pm$0.006 \\
2015-06-09 & 43 & J4 & 138.1$\pm$6.9 & 0.306$\pm$0.014 & 120.2$\pm$2.7 & 0.139$\pm$0.007 \\
2015-06-09 & 43 & J3 & 99.8$\pm$5.0 & 0.946$\pm$0.030 & 122.0$\pm$1.8 & 0.303$\pm$0.015 \\
2015-06-09 & 43 & J2 & 154.2$\pm$7.7 & 1.810$\pm$0.084 & 142.4$\pm$2.7 & 0.844$\pm$0.042 \\
2015-06-09 & 43 & J1 & 49.3$\pm$2.5 & 6.095$\pm$0.103 & 154.6$\pm$1.0 & 1.028$\pm$0.051 \\ \hline
2015-07-02 & 43 & C & 1278.8$\pm$63.9 & 0 & 0 & 0.027$\pm$0.001 \\
2015-07-02 & 43 & J5 & 481.5$\pm$24.1 & 0.145$\pm$0.011 & 116.9$\pm$4.6 & 0.110$\pm$0.006 \\
2015-07-02 & 43 & J4 & 106.5$\pm$5.3 & 0.312$\pm$0.012 & 117.7$\pm$2.3 & 0.121$\pm$0.006 \\
2015-07-02 & 43 & J3 & 79.5$\pm$4.0 & 0.954$\pm$0.025 & 121.7$\pm$1.5 & 0.248$\pm$0.012 \\
2015-07-02 & 43 & J2 & 143.5$\pm$7.2 & 1.751$\pm$0.088 & 141.3$\pm$2.9 & 0.877$\pm$0.044 \\
2015-07-02 & 43 & J1 & 38.7$\pm$1.9 & 6.016$\pm$0.057 & 154.7$\pm$0.5 & 0.572$\pm$0.029 \\ \hline
2015-08-01 & 43 & C & 1468.4$\pm$73.4 & 0 & 0 & 0.042$\pm$0.002 \\
2015-08-01 & 43 & J5 & 412.4$\pm$20.6 & 0.172$\pm$0.013 & 113.2$\pm$4.3 & 0.125$\pm$0.006 \\
2015-08-01 & 43 & J4 & 64.4$\pm$3.2 & 0.365$\pm$0.014 & 123.5$\pm$2.2 & 0.137$\pm$0.007 \\
2015-08-01 & 43 & J3 & 73.5$\pm$3.7 & 0.966$\pm$0.029 & 121.9$\pm$1.7 & 0.289$\pm$0.014 \\
2015-08-01 & 43 & J2 & 139.4$\pm$7.0 & 1.756$\pm$0.085 & 141.5$\pm$2.8 & 0.848$\pm$0.042 \\
2015-08-01 & 43 & J1 & 32.0$\pm$1.6 & 6.058$\pm$0.065 & 154.2$\pm$0.6 & 0.649$\pm$0.032 \\ \hline
2015-09-22 & 43 & C & 1674.0$\pm$83.7 & 0 & 0 & 0.043$\pm$0.002 \\
2015-09-22 & 43 & J5 & 399.4$\pm$20.0 & 0.188$\pm$0.015 & 112.7$\pm$4.6 & 0.155$\pm$0.008 \\
2015-09-22 & 43 & J4 & 40.2$\pm$2.0 & 0.384$\pm$0.009 & 128.7$\pm$1.3 & 0.087$\pm$0.004 \\
2015-09-22 & 43 & J3 & 91.8$\pm$4.6 & 0.974$\pm$0.044 & 121.6$\pm$2.6 & 0.442$\pm$0.022 \\
2015-09-22 & 43 & J2 & 142.2$\pm$7.1 & 1.765$\pm$0.081 & 140.1$\pm$2.6 & 0.815$\pm$0.041 \\
2015-09-22 & 43 & J1 & 63.4$\pm$3.2 & 6.133$\pm$0.139 & 155.2$\pm$1.3 & 1.392$\pm$0.070 \\ \hline
2015-12-05 & 43 & C & 1699.7$\pm$85.0 & 0 & 0 & 0.049$\pm$0.002 \\
2015-12-05 & 43 & J5 & 330.3$\pm$16.5 & 0.195$\pm$0.019 & 116.9$\pm$5.8 & 0.192$\pm$0.010 \\
2015-12-05 & 43 & J4 & 10.3$\pm$0.5 & 0.462$\pm$0.031 & 131.3$\pm$3.8 & 0.305$\pm$0.015 \\
2015-12-05 & 43 & J3 & 85.4$\pm$4.3 & 0.984$\pm$0.054 & 125.3$\pm$3.1 & 0.537$\pm$0.027 \\
2015-12-05 & 43 & J2 & 115.4$\pm$5.8 & 1.820$\pm$0.084 & 142.5$\pm$2.7 & 0.841$\pm$0.042 \\
2015-12-05 & 43 & J1 & 62.4$\pm$3.1 & 6.153$\pm$0.110 & 155.1$\pm$1.0 & 1.098$\pm$0.055 \\ \hline
2016-01-01 & 43 & C & 1839.3$\pm$92.0 & 0 & 0 & 0.049$\pm$0.002 \\
2016-01-01 & 43 & J5 & 458.5$\pm$22.9 & 0.169$\pm$0.022 & 121.0$\pm$5.0 & 0.219$\pm$0.011 \\
2016-01-01 & 43 & J4 & 12.2$\pm$0.6 & 0.519$\pm$0.027 & 130.7$\pm$2.6 & 0.271$\pm$0.014 \\
2016-01-01 & 43 & J3 & 111.9$\pm$5.6 & 1.006$\pm$0.057 & 126.6$\pm$3.0 & 0.568$\pm$0.028 \\
2016-01-01 & 43 & J2 & 132.7$\pm$6.6 & 1.868$\pm$0.081 & 142.5$\pm$2.4 & 0.805$\pm$0.040 \\
2016-01-01 & 43 & J1 & 73.2$\pm$3.7 & 6.136$\pm$0.109 & 155.5$\pm$1.0 & 1.088$\pm$0.054 \\ \hline
2016-01-25 & 15 & C & 1535.7$\pm$3.7 & 0 & 0 & 0.14$\pm$0.054 \\
2016-01-25 & 15 & K7 & 138.5$\pm$3.7 & 16.88$\pm$0.109 & 145.0$\pm$1.0 & 3.42$\pm$0.054 \\
2016-01-25 & 15 & K6 & 168.0$\pm$3.7 & 12.83$\pm$0.109 & 155.6$\pm$1.0 & 3.7$\pm$0.054 \\
2016-01-25 & 15 & K5 & 121.2$\pm$3.7 & 7.98$\pm$0.109 & 161.2$\pm$1.0 & 1.84$\pm$0.054 \\
2016-01-25 & 15 & K4 & 143.1$\pm$3.7 & 6.07$\pm$0.109 & 156.2$\pm$1.0 & 0.88$\pm$0.054 \\
2016-01-25 & 15 & K3 & 76.8$\pm$3.7 & 4.19$\pm$0.109 & 162.1$\pm$1.0 & 1.39$\pm$0.054 \\
2016-01-25 & 15 & K2 & 292.2$\pm$3.7 & 1.99$\pm$0.109 & 146.7$\pm$1.0 & 0.83$\pm$0.054 \\ 
2016-01-25 & 15 & K1 & 272.6$\pm$3.7 & 0.96$\pm$0.109 & 132.9$\pm$1.0 & 0.48$\pm$0.054 \\ \hline
2016-01-31 & 43 & C & 1643.6$\pm$82.2 & 0 & 0 & 0.057$\pm$0.003 \\
2016-01-31 & 43 & J5 & 304.6$\pm$15.2 & 0.171$\pm$0.020 & 119.0$\pm$7.0 & 0.204$\pm$0.010 \\
2016-01-31 & 43 & J4 & 47.7$\pm$2.4 & 0.612$\pm$0.030 & 130.7$\pm$2.8 & 0.296$\pm$0.015 \\
2016-01-31 & 43 & J3 & 28.2$\pm$1.4 & 1.172$\pm$0.025 & 124.2$\pm$1.2 & 0.252$\pm$0.013 \\
2016-01-31 & 43 & J2 & 126.0$\pm$6.3 & 1.702$\pm$0.083 & 142.1$\pm$2.8 & 0.834$\pm$0.042 \\
2016-01-31 & 43 & J1 & 37.4$\pm$1.9 & 6.031$\pm$0.052 & 154.5$\pm$0.5 & 0.524$\pm$0.026 \\ \hline
2016-03-18 & 43 & C & 1914.3$\pm$95.7 & 0 & 0 & 0.067$\pm$0.003 \\
2016-03-18 & 43 & J5 & 147.9$\pm$7.4 & 0.222$\pm$0.017 & 118.3$\pm$4.4 & 0.169$\pm$0.008 \\
2016-03-18 & 43 & J4 & 42.9$\pm$2.1 & 0.570$\pm$0.024 & 132.3$\pm$2.5 & 0.245$\pm$0.012 \\
2016-03-18 & 43 & J3 & 40.1$\pm$2.0 & 1.151$\pm$0.034 & 125.7$\pm$1.7 & 0.345$\pm$0.017 \\
2016-03-18 & 43 & J2 & 108.9$\pm$5.4 & 1.728$\pm$0.088 & 142.0$\pm$2.9 & 0.877$\pm$0.044 \\
2016-03-18 & 43 & J1 & 43.9$\pm$2.2 & 6.111$\pm$0.076 & 155.1$\pm$0.7 & 0.762$\pm$0.038 \\ \hline
2016-04-22 & 43 & C & 1963.2$\pm$98.2 & 0 & 0 & 0.061$\pm$0.003 \\
2016-04-22 & 43 & J5 & 173.3$\pm$8.7 & 0.248$\pm$0.022 & 126.9$\pm$5.2 & 0.224$\pm$0.011 \\
2016-04-22 & 43 & J4 & 39.3$\pm$2.0 & 0.665$\pm$0.026 & 131.4$\pm$2.3 & 0.263$\pm$0.013 \\
2016-04-22 & 43 & J3 & 31.6$\pm$1.6 & 1.258$\pm$0.030 & 126.8$\pm$1.4 & 0.299$\pm$0.015 \\
2016-04-22 & 43 & J2 & 102.6$\pm$5.1 & 1.727$\pm$0.092 & 142.8$\pm$3.0 & 0.919$\pm$0.046 \\
2016-04-22 & 43 & J1 & 21.4$\pm$1.1 & 6.143$\pm$0.056 & 153.3$\pm$0.5 & 0.559$\pm$0.028 \\ \hline
2016-06-10 & 43 & C & 2562.7$\pm$128.1 & 0 & 0 & 0.054$\pm$0.003 \\
2016-06-10 & 43 & J5 & 161.9$\pm$8.1 & 0.196$\pm$0.018 & 140.5$\pm$5.4 & 0.181$\pm$0.009 \\
2016-06-10 & 43 & J4 & 49.6$\pm$2.5 & 0.492$\pm$0.016 & 125.7$\pm$1.9 & 0.165$\pm$0.008 \\
2016-06-10 & 43 & J3 & 44.7$\pm$2.2 & 1.275$\pm$0.042 & 128.4$\pm$1.9 & 0.419$\pm$0.021 \\
2016-06-10 & 43 & J2 & 108.6$\pm$5.4 & 1.779$\pm$0.092 & 143.9$\pm$3.0 & 0.921$\pm$0.046 \\
2016-06-10 & 43 & J1 & 79.5$\pm$4.0 & 6.182$\pm$0.126 & 155.8$\pm$1.2 & 1.258$\pm$0.063 \\ \hline
2016-07-04 & 43 & C & 2580.7$\pm$129.0 & 0 & 0 & 0.049$\pm$0.002 \\
2016-07-04 & 43 & J5 & 142.0$\pm$7.1 & 0.317$\pm$0.024 & 127.4$\pm$4.4 & 0.245$\pm$0.012 \\
2016-07-04 & 43 & J4 & 31.5$\pm$1.6 & 0.759$\pm$0.029 & 128.6$\pm$2.2 & 0.288$\pm$0.014 \\
2016-07-04 & 43 & J3 & 27.3$\pm$1.4 & 1.346$\pm$0.025 & 129.4$\pm$1.1 & 0.248$\pm$0.012 \\
2016-07-04 & 43 & J2 & 117.7$\pm$5.9 & 1.739$\pm$0.098 & 142.9$\pm$3.2 & 0.984$\pm$0.049 \\
2016-07-04 & 43 & J1 & 59.3$\pm$3.0 & 6.169$\pm$0.105 & 155.2$\pm$1.0 & 1.053$\pm$0.053 \\ \hline
\end{longtable} 
\end{center}

\begin{table}
\centering
\caption{Component kinematics and jet properties from 43 GHz data.} \label{tab:ki}
\begin{tabular}{lll}
\hline \hline
Kinematic quantity & Symbol & Estimate \\ \hline
 Component proper motion (mas yr$^{-1}$)& J1 & 0.07 \\
  & J2 & 0.04 \\
  & J3 & 0.33 \\
  & J4 & 0.30 \\
  & J5 & 0.11 \\
  Apparent bulk speed (units of $c$) & $\beta_{\perp}$ & 17.5 \\
  Intrinsic bulk speed (units of $c$) & $\beta$ & $\geq 0.998$\\
  Bulk Lorentz factor & $\Gamma$ & $\geq 17.5$ \\
  Position angle & $\lambda$ & $128\fdg3$ \\
  Inclination angle & $i$ & $\leq 6\fdg6$ \\
  Projected half opening angle & $\psi$ & $15\fdg6$ \\
  Intrinsic half opening angle & $\theta_{0}$ & $\leq 1\fdg8$ \\ \hline
\end{tabular}
\end{table}

\end{document}